\documentclass[12pt]{article}

\usepackage[margin=1in]{geometry}
\usepackage[T1]{fontenc}
\usepackage[utf8]{inputenc}
\usepackage{mathptmx}
\usepackage{microtype}
\usepackage{amsmath,amssymb,amsthm,bm}
\usepackage{amsfonts}
\usepackage{booktabs,tabularx,array}
\usepackage{float}
\usepackage[dvipsnames]{xcolor}
\usepackage{enumitem}
\usepackage{natbib}
\usepackage{csquotes}
\usepackage{setspace}
\usepackage{fancyhdr}
\usepackage{titlesec}
\usepackage{caption}
\usepackage{graphicx}
\usepackage{subcaption}
\usepackage[colorlinks=true,linkcolor=MidnightBlue,citecolor=MidnightBlue,urlcolor=MidnightBlue]{hyperref}

\definecolor{statementblue}{HTML}{1F4E79}
\definecolor{lightblue}{HTML}{EAF1F7}

\titleformat{\section}{\large\bfseries\color{statementblue}}{\thesection.}{0.55em}{}
\titleformat{\subsection}{\normalsize\bfseries\color{statementblue}}{\thesubsection.}{0.55em}{}
\titleformat{\subsubsection}{\normalsize\itshape}{\thesubsubsection.}{0.55em}{}

\setlist{itemsep=0.2em,topsep=0.35em}
\title{\Huge\color{statementblue}\textbf{Structural Analysis of a Dynamic Multilayer Network via Matrix Autoregressive Models: A Case Study of International Interactions between Countries}}

\author{
\textbf{C. Pinzón\footnote{Corresponding author: \href{mailto:cpinzonca@unal.edu.co}{cpinzonca@unal.edu.co}.}\qquad M. Arrieta\qquad J. Sosa}\\
Departamento de Estadística, Universidad Nacional de Colombia\\
Bogotá D.C., Colombia
}

\date{}

\begin{document}

\maketitle

\begin{abstract}
Dynamic and multilayer networks have been widely studied separately, but their joint analysis remains comparatively underdeveloped. Because the relational information of a dynamic multilayer network can be represented as a tensor at each time point $t$, each layer can be summarized through a set of structural statistics, yielding a matrix-valued observation and, consequently, a matrix-valued time series. To exploit this structure, we propose the use of matrix autoregressive (MAR) models, which simultaneously characterize temporal dependence across relational layers and structural statistics. We apply this framework to the ICEWS dataset, which records international interactions among countries under four relational domains and therefore naturally defines a dynamic multilayer network. The results indicate that negative verbal interactions (\textit{Verbal-}) play a prominent role in the subsequent structural reconfiguration of the material-interaction layers, while mean strength exhibits the strongest temporal persistence and reciprocity the broadest cross-statistic influence. These findings illustrate the usefulness of MAR models for providing a parsimonious and interpretable characterization of temporal and cross-layer dependence in dynamic multilayer networks.
\end{abstract}


\newpage

\section{Introduction}

Relational systems are often both dynamic and multilayered. Interactions evolve over time while occurring through different types of relations, so their structure depends jointly on temporal ordering, cross-layer dependence, and network topology. Temporal-network methods emphasize the evolution and persistence of ties \citep{HolmeSaramaki2012}, whereas multilayer-network methods account for interactions across multiple relational channels \citep{Kivela2014,Boccaletti2014}. Although these two dimensions naturally coexist in applications such as communication, finance, transportation, and international relations, their joint structural evolution remains comparatively less studied.

Several statistical approaches address particular aspects of this problem. Temporal exponential random graph models describe the evolution of network ties through endogenous dependence mechanisms \citep{Hanneke2010,KrivitskyHandcock2014}, while dynamic latent-space and stochastic block models characterize changes in latent geometry or community structure \citep{SewellChen2015,MatiasMiele2017}. Multislice formulations accommodate temporal and multiplex structure, particularly for community detection \citep{Mucha2010}. Tensor-based approaches provide a complementary perspective. For example, \cite{Hoff2015} models longitudinal relational arrays through multilinear tensor regression, whereas factor network autoregressions use network structure to explain the temporal behavior of nodal variables \citep{FNAR}. These methods are primarily designed to model individual ties, latent relational structure, communities, or nodal outcomes. They do not directly target the temporal propagation of the \emph{global structural configuration} of a multilayer network.

This distinction defines the focus of the present work. Rather than modeling individual edges, we characterize each layer at every time point through a set of global network statistics describing complementary aspects of topology and interaction intensity. Stacking these statistics across layers produces a matrix-valued observation and transforms the evolving multilayer network into a matrix-valued time series. We then model this structural-state process using matrix autoregressive (MAR) models \citep{MAR}. Their bilinear specification is particularly attractive in this context because the two autoregressive coefficient matrices separately characterize temporal dependence across network layers and across structural statistics. This representation is substantially more parsimonious and interpretable than an unrestricted vector autoregression of all layer--statistic combinations, while impulse-response functions describe how structural perturbations propagate throughout the system.

The proposed framework therefore provides a descriptive and exploratory approach to the macroscopic dynamics of multilayer networks. Its main contribution is the integration of network structural summaries with matrix-valued time-series modeling, allowing the analyst to identify which layers drive subsequent structural change, which network characteristics exhibit temporal carryover, and how shocks to specific layer--statistic combinations propagate over time. We additionally assess whether the bilinear MAR representation remains informative when the underlying network dynamics are nonlinear. To this end, we construct a simulation mechanism combining autoregressive edge dynamics, thresholding, triangulation, and partial reciprocity while explicitly controlling the true cross-layer coupling structure. This provides a direct assessment of the ability of the MAR model to recover structural dependence from network processes that do not themselves follow a bilinear model.

We illustrate the methodology using the Integrated Crisis Early Warning System (ICEWS) data, which record international interactions classified into negative material, positive material, negative verbal, and positive verbal relations. Following \cite{Hoff2015}, we focus on a reduced set of highly relevant countries and obtain monthly multilayer networks for $25$ countries over $T=132$ time points. Each layer is summarized using density, assortativity, transitivity, reciprocity, average geodesic distance, and mean strength, producing the matrix-valued series used for MAR estimation. The empirical results reveal substantial asymmetry in the system. Negative verbal interactions emerge as an important source of cross-layer structural propagation, affecting the subsequent configuration of both material-interaction layers, whereas their own structure is not significantly explained by the other layers. Mean strength exhibits the strongest temporal persistence and structural carryover. Impulse-response analysis further shows that shocks to verbal-conflict density generate broad but short-lived responses, whereas shocks to the mean strength of material conflict produce more persistent effects and are followed by a short-run deterioration in verbal cooperation. These findings are consistent with the empirical patterns reported throughout the analysis.

The simulation study provides a complementary assessment of the proposed representation. Across nine scenarios, the MAR model generally achieves high sensitivity for detecting genuine cross-layer dependencies, although performance deteriorates when the time series is shorter or the network becomes larger. Specificity is notably lower, with median values around $0.7$, indicating that some nonexistent dependencies may be declared significant. Thus, the bilinear approximation captures dominant channels of structural propagation even when the underlying network-generating mechanism is nonlinear, but its estimated dependence structure should be interpreted primarily as an exploratory description rather than as exact support recovery.

The remainder of the paper is organized as follows. Section~2 develops the structural-state representation and reviews matrix autoregressive dynamics. Section~3 presents the ICEWS application, including data construction, model estimation, impulse-response analysis, and diagnostics. Section~4 evaluates the proposed approach through a simulation study. Section~5 concludes with a discussion of the main findings, limitations, and directions for future research.


\section{Structural-State Dynamics of Multilayer Networks}

Our analysis represents a dynamic multilayer network through the temporal evolution of a small set of global structural characteristics and models the resulting matrix-valued series using matrix autoregression. This construction separates dependence across relational layers from dependence across network characteristics, while preserving a parsimonious representation suitable for the moderate time dimension of the application. We introduce below only the concepts required for the empirical analysis.

\subsection{Structural-state representation}

Consider a dynamic multilayer network observed at times $t=1,\ldots,T$, with a common set of $n$ actors and $m$ relational layers. Let $\mathbf{W}_{t}^{(r)}=(w_{ij,t}^{(r)})$ denote the weighted adjacency matrix of layer $r$ at time $t$, where $w_{ij,t}^{(r)}$ measures the intensity of the directed interaction from actor $i$ to actor $j$. The collection $\{\mathbf{W}_{t}^{(r)}:r=1,\ldots,m\}$ therefore represents the complete multilayer network at time $t$.

Rather than modeling individual dyads, our inferential target is the \emph{global structural state} of the system. Each layer is therefore summarized by $l$ network statistics and represented by the vector $\mathbf{s}_{t}^{(r)}=(s_{1,t}^{(r)},\ldots,s_{l,t}^{(r)})^\top$. Stacking these vectors across layers yields the matrix-valued time series
\[
\mathbf{X}_t
=
\begin{pmatrix}
\mathbf{s}_{t}^{(1)\top}\\
\vdots\\
\mathbf{s}_{t}^{(m)\top}
\end{pmatrix}
\in\mathbb{R}^{m\times l},
\qquad t=1,\ldots,T.
\]
Thus, rows correspond to relational layers and columns to structural characteristics. This reduction sacrifices dyad-specific information in exchange for a parsimonious description of the macroscopic organization of the multilayer system. It is appropriate when the objective is to study how global network structure propagates across layers and over time rather than to explain the formation or dissolution of individual ties.

We characterize each layer through density, assortativity, transitivity, reciprocity, average geodesic distance, and mean strength, which capture complementary dimensions of connectivity, mixing, local closure, mutuality, separation, and interaction intensity \citep{libroguia,Newman2018}. Density measures the prevalence of realized ties among those that are possible; assortativity characterizes the tendency of actors with similar connectivity profiles to interact; transitivity quantifies local closure; reciprocity measures the extent to which directed interactions are reciprocated; average geodesic distance describes typical network separation; and mean strength summarizes the average intensity of weighted interactions. Because some of these quantities admit different conventions for directed, weighted, and disconnected networks, the computational definitions used in the empirical analysis are specified in Section~3 and applied consistently across layers and time points.

\subsection{Temporal preprocessing and stationarity}

The MAR framework assumes a stationary matrix-valued process \citep{MAR,Review}. Global network statistics, however, may display long-run changes that reflect gradual evolution of the relational system rather than the short-run dependence of primary interest. For a generic structural series $z_t$, an additive decomposition may be written as $z_t=\mu_t+S_t+a_t$, where $\mu_t$ represents trend, $S_t$ a deterministic seasonal component, and $a_t$ the remaining stochastic variation \citep{Daniel}. In the absence of relevant seasonality, the analysis reduces to separating a smooth temporal component from short-run fluctuations.

Accordingly, each layer--statistic series is detrended before MAR estimation. In the empirical application, the smooth component is estimated using locally weighted regression \citep{Cleveland1979}. The detrended series are subsequently standardized to zero mean and unit variance because the network statistics are measured on markedly different scales. This step prevents scale differences from dominating least-squares estimation and facilitates comparison of temporal effects across structural characteristics.

Stationarity of the resulting marginal series is assessed using the Augmented Dickey--Fuller test \citep{ADF}. Such marginal tests provide evidence against unit-root behavior but do not, by themselves, establish joint stationarity of the matrix-valued process. Model adequacy is therefore additionally evaluated through the stability of the fitted MAR process and residual diagnostics.

\subsection{Matrix autoregressive dynamics}

Vector autoregressive models provide a standard representation of temporal dependence among jointly observed variables \citep{VAR,VAR2,Lutkepohl2005}. Directly vectorizing $\mathbf{X}_t\in\mathbb{R}^{m\times l}$ and fitting an unrestricted VAR, however, ignores its two-dimensional organization and requires $(ml)^2$ autoregressive coefficients at each lag. Matrix autoregressive models preserve this structure by imposing a bilinear restriction on temporal dependence \citep{MAR,Review}.

For a first-order model,
\[
\mathbf{X}_t
=
\mathbf{A}_1\mathbf{X}_{t-1}\mathbf{A}_2^\top
+
\mathbf{E}_t,
\]
where $\mathbf{A}_1\in\mathbb{R}^{m\times m}$ captures dependence across layers, $\mathbf{A}_2\in\mathbb{R}^{l\times l}$ captures dependence across structural statistics, and $\mathbf{E}_t$ is a zero-mean matrix innovation. Under the standard column-wise vectorization convention, this model is equivalent to
\[
\operatorname{vec}(\mathbf{X}_t)
=
(\mathbf{A}_2\otimes\mathbf{A}_1)
\operatorname{vec}(\mathbf{X}_{t-1})
+
\operatorname{vec}(\mathbf{E}_t),
\]
so the MAR(1) model is a structured VAR(1) whose autoregressive matrix is constrained to have Kronecker-product form \citep{MAR}.

This restriction provides substantial dimension reduction. An unrestricted VAR(1) contains $m^2l^2$ autoregressive coefficients, whereas the MAR(1) representation contains only $m^2+l^2$ before identification constraints. In our application, with $m=4$ layers and $l=6$ structural statistics, this corresponds to $576$ coefficients for an unrestricted VAR compared with $52$ coefficients in the MAR representation.

The bilinear structure also provides the interpretation required for the empirical analysis. An entry of $\mathbf{A}_1$ describes how the previous structural profile of one layer contributes to the subsequent profile of another layer, whereas an entry of $\mathbf{A}_2$ describes temporal propagation from one structural characteristic to another under a common cross-layer mechanism. The two matrices therefore distinguish \emph{cross-layer} from \emph{cross-statistic} dependence, which is the central inferential advantage of the representation considered here.

Because $\rho(\mathbf{A}_2\otimes\mathbf{A}_1)=\rho(\mathbf{A}_2)\rho(\mathbf{A}_1)$, with $\rho(\cdot)$ representing the spectral radius operator, the MAR(1) process is stable when $\rho(\mathbf{A}_1)\rho(\mathbf{A}_2)<1$ \citep{MAR}. The bilinear parameterization is not uniquely identified with respect to scale and sign because multiplying $\mathbf{A}_1$ by a nonzero constant and $\mathbf{A}_2$ by its reciprocal leaves the fitted process unchanged. We resolve the scale indeterminacy by imposing $\lVert\mathbf{A}_1\rVert_F=1$ and the remaining sign indeterminacy by requiring $\operatorname{tr}(\mathbf{A}_1)>0$. These identification conventions are used throughout estimation and interpretation.

Higher-order temporal dependence is accommodated naturally by replacing the first-order conditional mean with $\sum_{h=1}^{p}\mathbf{A}_{1,h}\mathbf{X}_{t-h}\mathbf{A}_{2,h}^\top$. This formulation motivates the comparison of alternative lag orders carried out in Section~3.

\subsubsection{Estimation}

The MAR parameters can be estimated by least squares. For the first-order model, the estimator minimizes
\[
Q(\mathbf{A}_1,\mathbf{A}_2)
=
\sum_{t=2}^{T}
\left\|
\mathbf{X}_t
-
\mathbf{A}_1\mathbf{X}_{t-1}\mathbf{A}_2^\top
\right\|_F^2
\]
subject to the identification constraints. Because the criterion is bilinear rather than jointly linear in $\mathbf{A}_1$ and $\mathbf{A}_2$, estimation can be implemented through alternating least squares, updating one coefficient matrix conditional on the other. A useful initialization is obtained by projecting the unrestricted VAR estimate onto the space of Kronecker-product coefficient matrices. Under the regularity conditions established by \cite{MAR}, the resulting least-squares estimators are consistent and $\sqrt{T}$-asymptotically normal, providing the basis for statistical inference on the individual coefficients.

\subsubsection{Impulse-response analysis}

Coefficient matrices describe one-step dependence, but they do not directly reveal how a perturbation propagates through the system over several periods. Because the vectorized MAR(1) model is a structured VAR with transition matrix ${\Phi}=\mathbf{A}_2\otimes\mathbf{A}_1$, standard impulse-response analysis applies \citep{Lutkepohl2005}. A one-time innovation $\mathbf{u}$ has horizon-$h$ effect
\[
\operatorname{IRF}(h;\mathbf{u})
=
{\Phi}^{h}\mathbf{u},
\qquad h=0,1,\ldots,
\]
which converges to zero under stability.

When contemporaneous innovations are correlated, orthogonalized impulse responses separate shocks by factorizing the innovation covariance matrix \citep{Lutkepohl2005}. These responses quantify the magnitude, direction, and persistence with which a perturbation originating in a particular layer--statistic combination propagates to the remaining components of the structural state. When orthogonalization is based on a Cholesky decomposition, the resulting responses depend on the ordering of the variables; the ordering used in the empirical analysis must therefore be fixed and reported explicitly.


\section{Structural Dynamics of International Interactions}

We apply the structural-state framework developed in Section~2 to the Integrated Crisis Early Warning System (ICEWS) data. The objective is to characterize the short-run evolution of the global topology of international interactions, with particular emphasis on three questions: which relational layers are most strongly associated with subsequent structural reconfiguration, which network characteristics transmit temporal dependence, and how structural innovations propagate through the multilayer system. The analysis proceeds by constructing monthly multilayer networks, mapping them into the matrix-valued structural-state representation, preprocessing the resulting series, selecting and estimating an appropriate MAR specification, and evaluating its dynamics through mode-specific coefficient matrices, impulse-response functions, and residual diagnostics.

\subsection{Data and structural-state construction}

The ICEWS data are publicly available through the Harvard Dataverse at \url{https://dataverse.harvard.edu/dataverse/icews}. They contain daily counts of directed interactions among $248$ countries from early 2004 to mid-2014, classified according to the CAMEO event taxonomy \citep{gerner2002creation}. Following the relational aggregation considered by \citet{Hoff2015}, interactions are grouped into four substantively distinct layers: negative material (\textit{Material-}), positive material (\textit{Material+}), negative verbal (\textit{Verbal-}), and positive verbal (\textit{Verbal+}) interactions. The resulting system is therefore a dynamic multilayer network with $m=4$ relational layers.

Daily interactions are aggregated by month. To reduce the dimensionality of the relational system while retaining its most structurally prominent actors, we restrict the analysis to the $25$ countries with the largest PageRank centrality averaged across both time and layers. This selection is stable across relational domains: when PageRank is averaged over time separately within each layer, $19$ countries belong to the top $25$ in all four rankings. The final set therefore represents a persistent core of highly central actors in the international-interaction system.\footnote{The selected actors are the United States, Israel, Russia, China, Palestine, Iran, Iraq, United Kingdom, Afghanistan, Pakistani Territory, India, Syria, Japan, France, South Korea, Turkey, North Korea, Australia, Lebanon, Somalia, Sudan, Egypt, Ukraine, Germany, and Serbia. \citet{Hoff2015} includes Georgia and Taiwan rather than Somalia and Serbia.}

With $n=25$ countries, $m=4$ layers, and $T=132$ monthly observations, the complete relational data are represented by the array $\mathcal{Y}=(y_{i_1,i_2,r,t})$, where $y_{i_1,i_2,r,t}$ is the number of interactions of type $r$ initiated by country $i_1$ toward country $i_2$ during month $t$. Rather than modeling the $n\times n\times m$ relational array directly, we construct the structural-state representation introduced in Section~2. At each time point, every layer is summarized by density, assortativity, transitivity, reciprocity, average geodesic distance, and mean strength. The observation at time $t$ is therefore a $4\times6$ matrix $\mathbf{X}_t$, whose rows correspond to relational layers and whose columns correspond to structural characteristics.

For the empirical construction of the structural state, density, assortativity, transitivity, reciprocity, and average geodesic distance are computed from the binary support of each weighted adjacency matrix, with an edge defined by $\mathbb{I}(w_{ij,t}^{(r)}>0)$, whereas mean strength is computed from the original interaction counts and therefore retains information on tie intensity. Assortativity is based on node degree, transitivity measures triadic closure, and reciprocity summarizes the prevalence of mutual directed ties. Average geodesic distance is computed using shortest-path distances among connected pairs. Mean strength is defined from total node strength, combining incoming and outgoing interaction weights. The same definitions are used for every layer and time point to ensure comparability of the resulting structural-state series.

\begin{figure}[!b]
    \centering
    \begin{subfigure}[t]{0.48\textwidth}
        \centering
        \includegraphics[width=\linewidth]{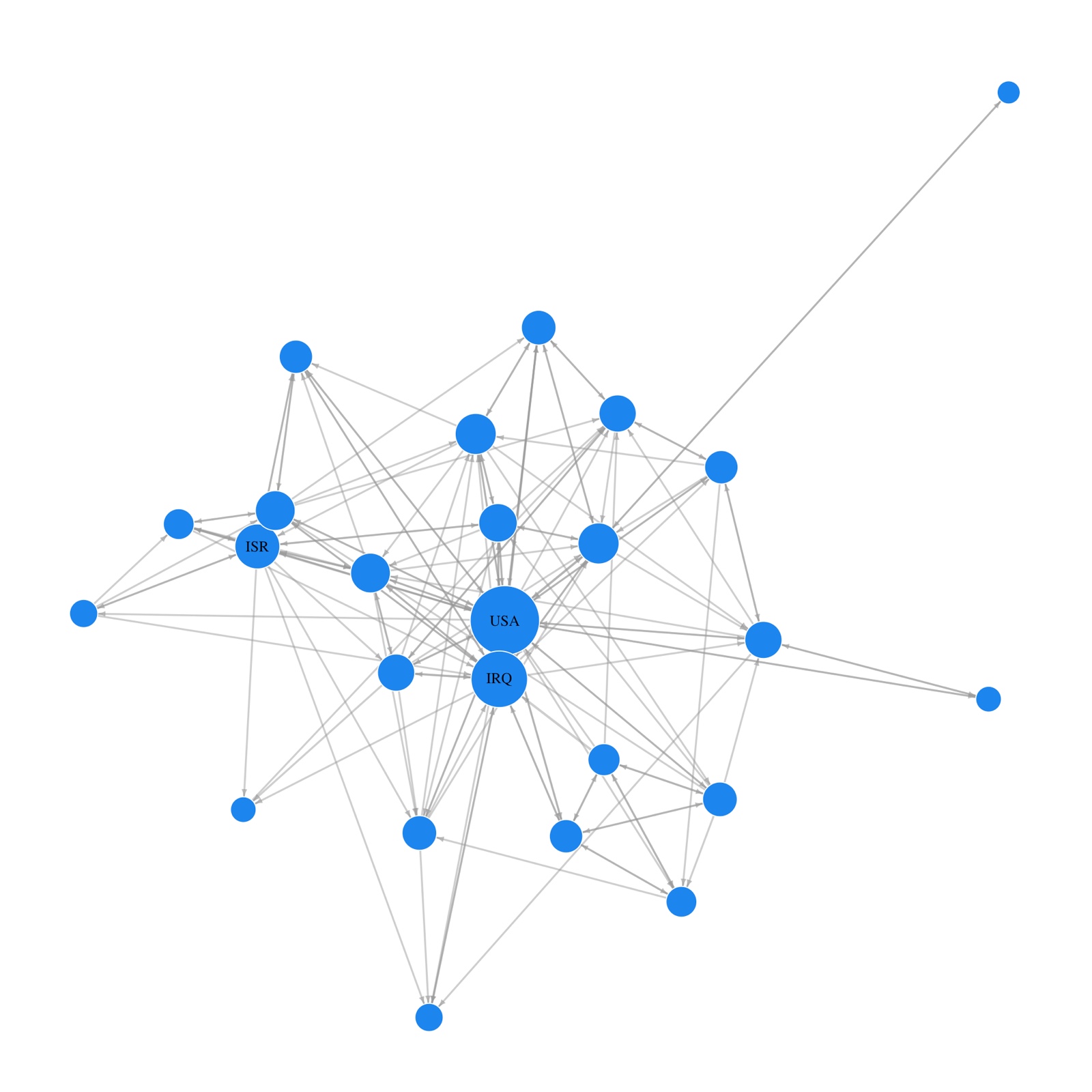}
        \caption{\textit{Material-} layer at $t=1$.}
        \label{fig:red_material_ini}
    \end{subfigure}
    \hfill
    \begin{subfigure}[t]{0.48\textwidth}
        \centering
        \includegraphics[width=\linewidth]{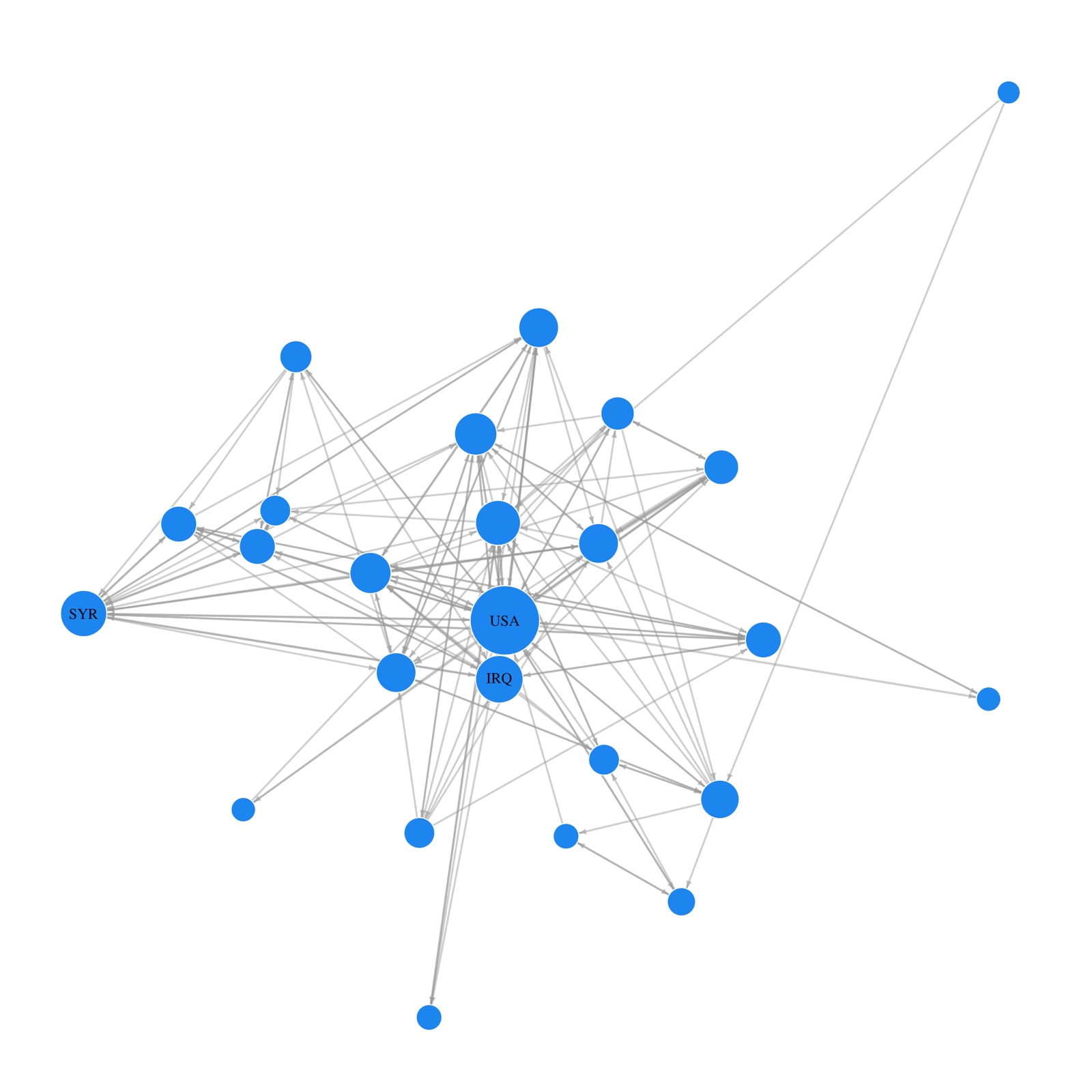}
        \caption{\textit{Material-} layer at $t=132$.}
        \label{fig:red_material_fin}
    \end{subfigure}

    \vspace{0.6em}

    \begin{subfigure}[t]{0.48\textwidth}
        \centering
        \includegraphics[width=\linewidth]{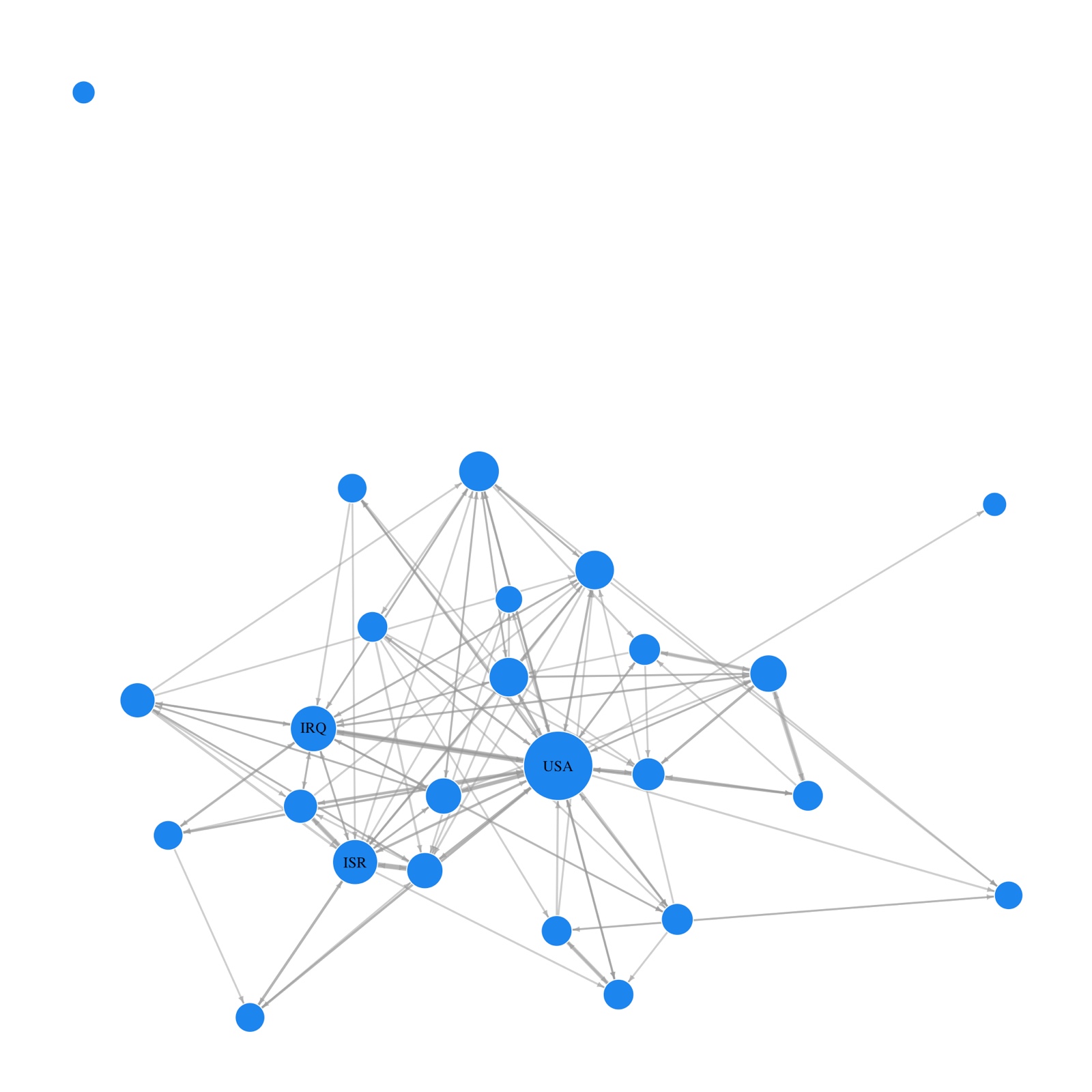}
        \caption{\textit{Verbal-} layer at $t=1$.}
        \label{fig:red_verbal_ini}
    \end{subfigure}
    \hfill
    \begin{subfigure}[t]{0.48\textwidth}
        \centering
        \includegraphics[width=\linewidth]{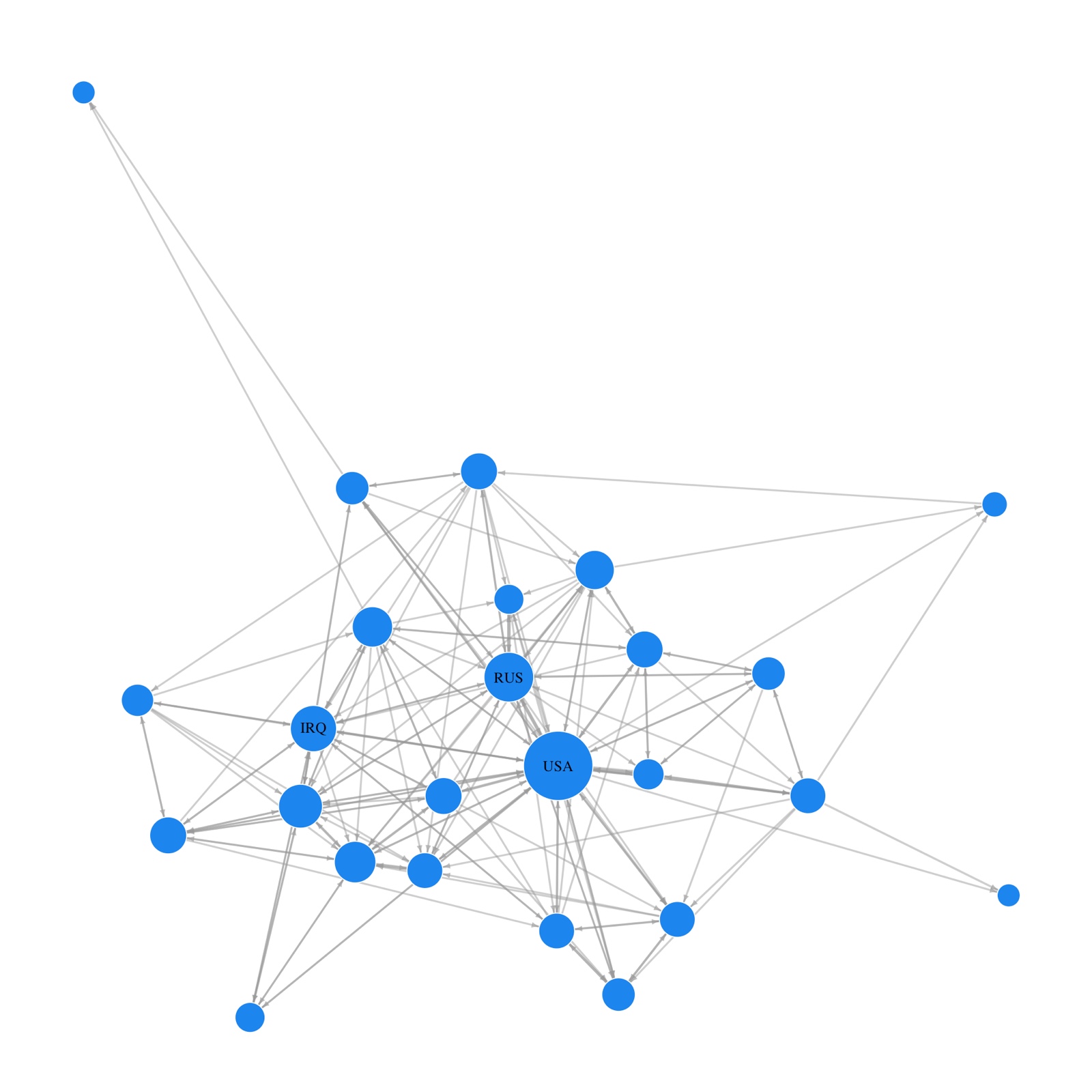}
        \caption{\textit{Verbal-} layer at $t=132$.}
        \label{fig:red_verbal_fin}
    \end{subfigure}

    \caption{Snapshots of the \textit{Material-} and \textit{Verbal-} networks at the initial and final time points. Node size is proportional to total degree, and the three highest-degree countries are labeled.}
    \label{fig:redes_ejemplo}
\end{figure}

To complement the structural-state representation, Figure~\ref{fig:redes_ejemplo} shows the \textit{Material-} and \textit{Verbal-} networks at $t=1$ and $t=132$. Node size is proportional to total degree, and the three highest-degree countries are labeled. The snapshots reveal both persistence and temporal reconfiguration: the United States and Iraq remain among the most connected actors in both layers, while the third leading country changes over time. Israel is prominent at $t=1$ in both layers, whereas Syria emerges in \textit{Material-} and Russia in \textit{Verbal-} at $t=132$. This pattern illustrates the coexistence of a persistent structural core with meaningful changes in the distribution of connectivity.

\begin{figure}[!b]
    \centering
    \includegraphics[width=\linewidth]{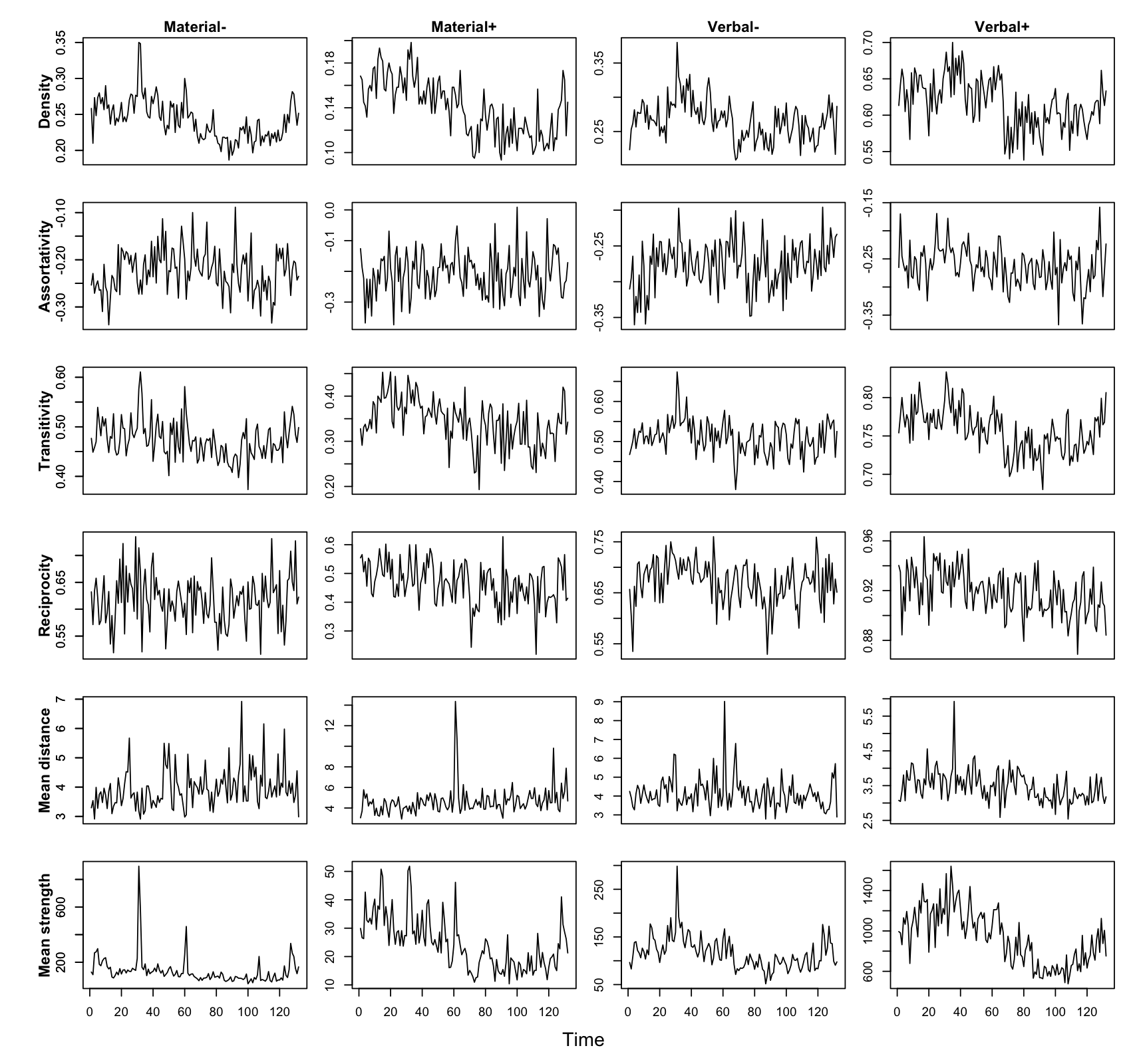}
    \caption{Monthly trajectories of the six structural characteristics for the four international-interaction layers before temporal preprocessing.}
    \label{fig:estadisticas}
\end{figure}

The six statistics were selected to capture complementary dimensions of network organization while avoiding redundancy and temporally uninformative summaries. The number of connected components and clique number, for example, were excluded because at least one corresponding layer-specific series was constant. Average geodesic distance was preferred to diameter because it summarizes separation across the entire network rather than depending on a single extreme pair of actors. Mean strength was preferred to mean degree because, with a fixed number of actors, mean degree is directly related to density and would therefore introduce substantial redundancy into the structural-state representation.

\begin{figure}[!b]
    \centering
    \includegraphics[width=\linewidth]{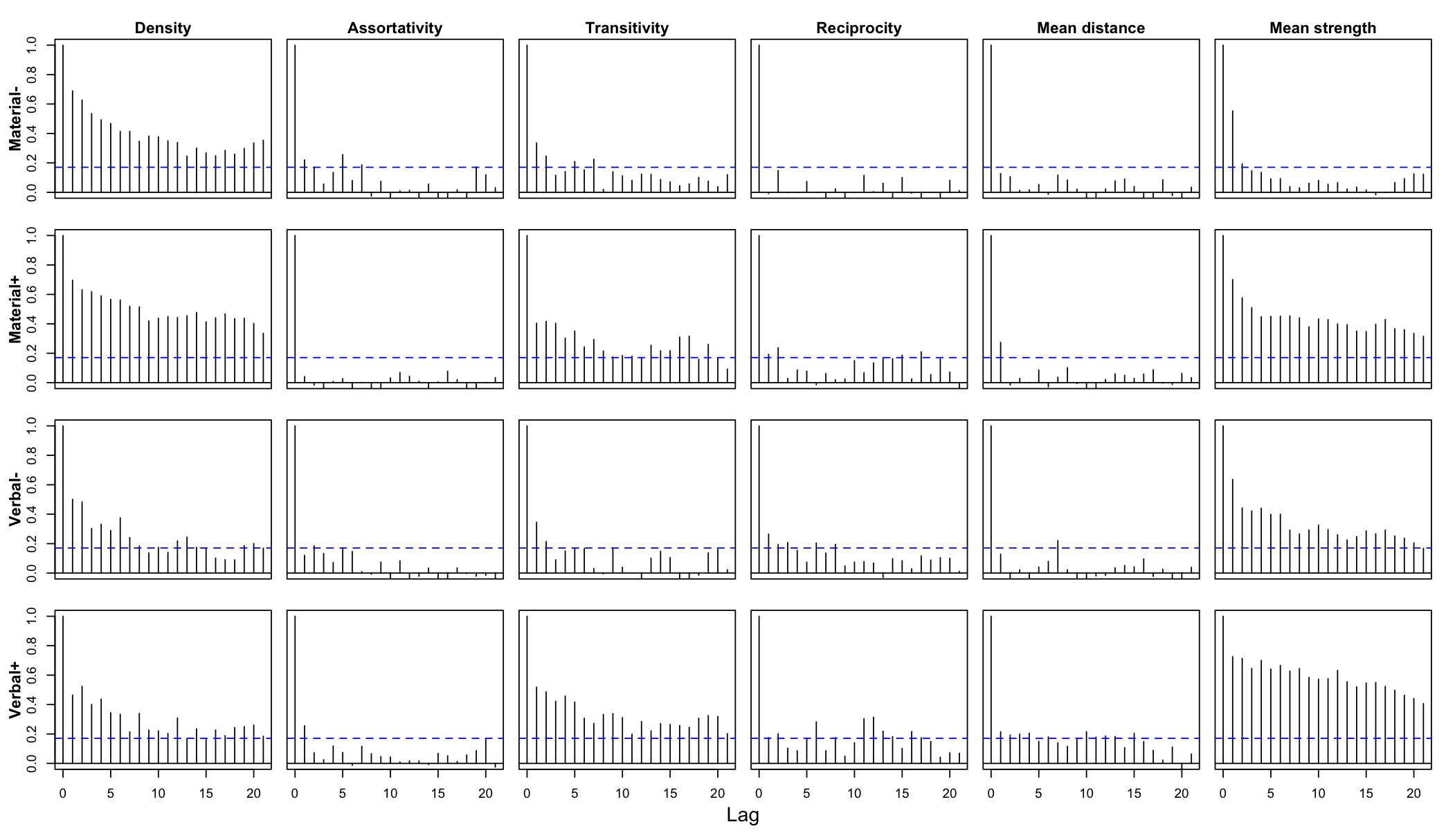}
    \caption{Autocorrelation functions for the structural-state series before detrending and standardization.}
    \label{fig:acf}
\end{figure}

Figure~\ref{fig:estadisticas} shows the monthly trajectories of the resulting $24$ layer--statistic series. Several systematic differences across relational layers are apparent. The \textit{Verbal+} layer is consistently the densest and also exhibits comparatively high levels of transitivity and reciprocity, indicating a more connected and locally closed relational structure. Assortativity remains close to zero across the four layers, providing little evidence of persistent degree-based assortative mixing. The density trajectories also exhibit a visible change in their evolution around month $68$. A particularly pronounced episode occurs around month $31$, approximately in mid-2006, when mean strength in the \textit{Material-} layer increases sharply, indicating an unusually large concentration of negative material interactions during that period.

\begin{figure}[!b]
    \centering
    \includegraphics[width=\linewidth]{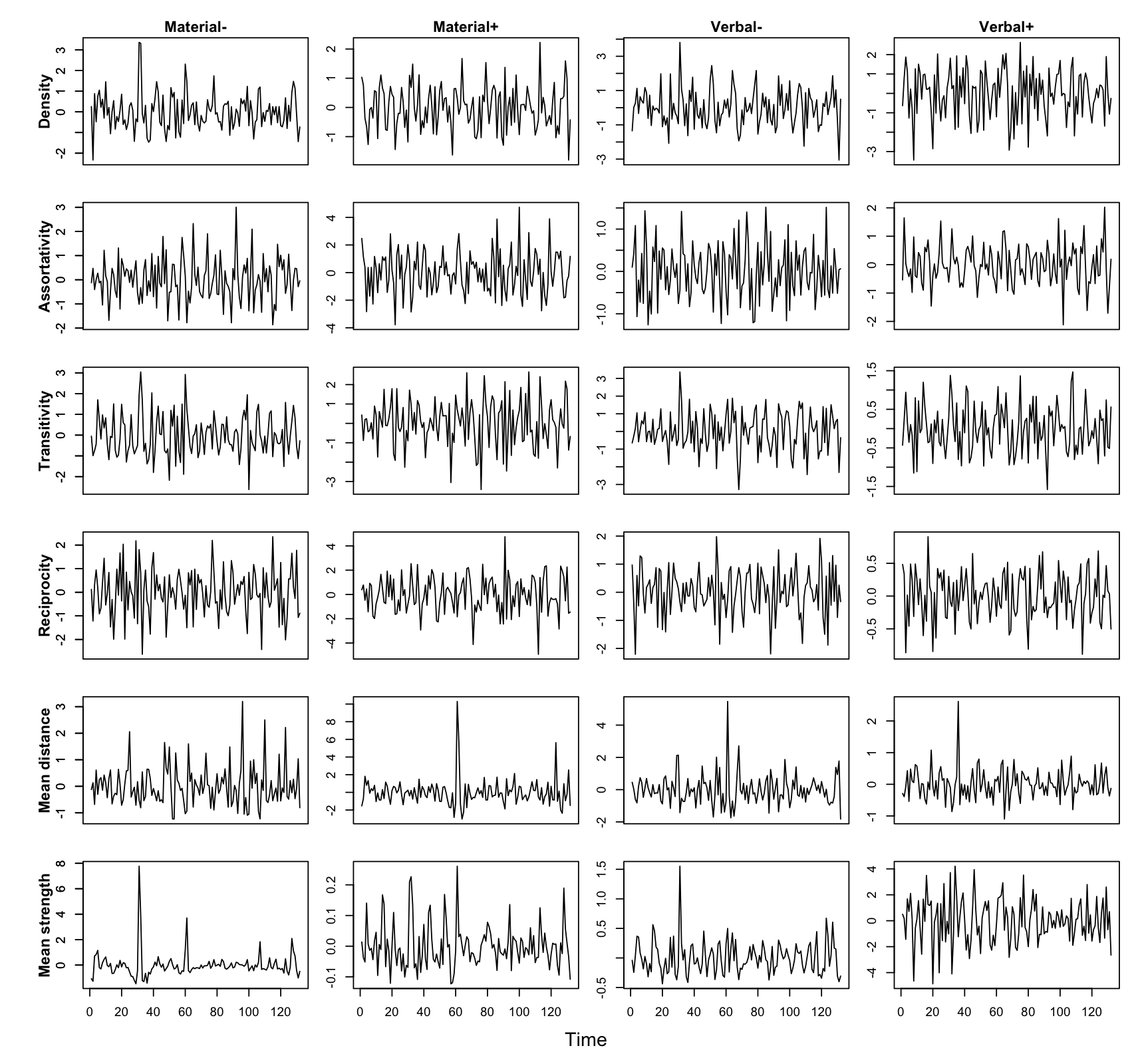}
    \caption{Standardized and detrended structural-state series used for MAR estimation.}
    \label{fig:estadisticas_final}
\end{figure}

Beyond these cross-layer differences, the raw structural series display substantial low-frequency temporal variation. Figure~\ref{fig:acf} shows particularly persistent autocorrelation for density, transitivity, and mean strength, with correlations that decay slowly across lags. This behavior is consistent with the presence of smooth temporal evolution that should be removed before the short-run autoregressive dependence is modeled.

Following Section~2, a smooth trend is estimated separately for each layer--statistic series using LOESS \citep{Cleveland1979} and removed from the observed trajectory. The detrended series are subsequently standardized to zero mean and unit variance. This preprocessing separates gradual structural evolution from the shorter-run dependence targeted by the MAR model and places the six statistics on comparable scales. Figure~\ref{fig:estadisticas_final} displays the matrix-valued series used for estimation.

Marginal stationarity is assessed using Augmented Dickey--Fuller tests \citep{ADF}. The unit-root null hypothesis is rejected at the nominal $5\%$ level for all $24$ layer--statistic series. These results do not by themselves establish joint stationarity of the matrix-valued process, but, together with the stability and residual assessments reported below, they support the use of a stationary MAR specification for the short-run dynamics.

\subsection{Model selection and mode-specific dependence}

MAR models with autoregressive orders $p=1,\ldots,6$ are estimated by least squares and compared using the Bayesian information criterion. The MAR(1) specification achieves the smallest BIC and is therefore retained for the empirical analysis. After removal of smooth long-run variation, the dominant temporal dependence in the structural-state process is thus adequately represented by the immediately preceding monthly state without requiring additional lags. The reported estimates use the identification constraints $\lVert\widehat{\mathbf{A}}_1\rVert_F=1$ and $\operatorname{tr}(\widehat{\mathbf{A}}_1)>0$, which fix the scale and sign of the bilinear decomposition.

Under the orientation defined in Section~2, $\widehat{\mathbf{A}}_1$ describes dependence along the layer mode and $\widehat{\mathbf{A}}_2$ along the structural-statistic mode. These matrices should be interpreted jointly rather than as two independent sets of conventional regression coefficients. Specifically, the implied one-step coefficient relating statistic $q$ in layer $j$ at time $t-1$ to statistic $k$ in layer $i$ at time $t$ is $\widehat{a}_{1,ij}\widehat{a}_{2,kq}$. Consequently, $\widehat{\mathbf{A}}_1$ identifies the dominant channels of cross-layer propagation, whereas $\widehat{\mathbf{A}}_2$ identifies the structural characteristics through which that propagation operates. This distinction is essential for interpreting the separable MAR representation.

Table~\ref{tab:A1} reports the estimated layer-mode coupling matrix. The most salient feature of $\widehat{\mathbf{A}}_1$ is the marked asymmetry of the cross-layer dynamics. Both material layers exhibit positive own-layer persistence, with coefficients $0.407$ for \textit{Material-} and $0.359$ for \textit{Material+}. Material cooperation is also associated with the subsequent configuration of verbal cooperation through an estimated layer-mode coefficient of $0.186$.

\begin{table}[!htb]
    \centering
    \small
    \begin{tabular}{lrrrr}
        \toprule
        & Material- & Material+ & Verbal- & Verbal+ \\
        \midrule
        Material- & \textbf{0.407} & -0.079 & \textbf{-0.249} & -0.019 \\
        Material+ & -0.012 & \textbf{0.359} & \textbf{-0.723} & 0.009 \\
        Verbal-   & 0.107 & 0.062 & 0.008 & -0.007 \\
        Verbal+   & -0.157 & \textbf{0.186} & 0.195 & -0.018 \\
        \bottomrule
    \end{tabular}
    \caption{Estimated layer-mode coupling matrix $\widehat{\mathbf{A}}_1$ for the MAR(1) model. Columns correspond to source layers at time $t-1$ and rows to recipient layers at time $t$. Bold entries are statistically distinguishable from zero at the nominal $5\%$ level.}
    \label{tab:A1}
\end{table}

More importantly, the column corresponding to \textit{Verbal-} contains the two strongest statistically distinguishable off-diagonal couplings in absolute magnitude. Its connection with the subsequent \textit{Material-} configuration is $-0.249$, while the corresponding coefficient for \textit{Material+} is $-0.723$, the largest entry in absolute value in the estimated layer-mode matrix. Conversely, none of the coefficients in the \textit{Verbal-} row is statistically distinguishable from zero. Thus, within the separable MAR representation, negative verbal interactions occupy a distinctive position: their lagged structural configuration is strongly connected to subsequent changes in the material layers, whereas the structural state of verbal conflict itself exhibits comparatively little dependence on the preceding layer configuration.

This source--recipient asymmetry is one of the central empirical findings. It does not imply that verbal conflict causally generates material conflict or cooperation, nor should the sign of an entry of $\widehat{\mathbf{A}}_1$ be interpreted independently of $\widehat{\mathbf{A}}_2$. Rather, it shows that the \textit{Verbal-} layer constitutes a dominant channel in the estimated layer-mode dependence. In contrast, the entire \textit{Verbal+} source column contains no statistically distinguishable entries, suggesting a substantially weaker role for positive verbal interactions in transmitting short-run structural variation to other relational domains.

The complementary statistic-mode structure is reported in Table~\ref{tab:A2}. Two complementary mechanisms dominate the statistic-mode dynamics. Reciprocity has the broadest pattern of statistically distinguishable cross-statistic couplings. Its lagged column contains negative coefficients for density, assortativity, and reciprocity itself, together with a positive coefficient for average geodesic distance. Within the common statistic-mode transformation, episodes characterized by unusually high reciprocity are therefore associated with subsequent movement toward lower connectivity and assortative mixing and greater network separation. These coefficients should again be interpreted jointly with the relevant entries of $\widehat{\mathbf{A}}_1$, because the sign and magnitude of a complete layer-specific transition are determined by their product.

\begin{table}[!htb]
    \centering
    \small
    \setlength{\tabcolsep}{4pt}
    \begin{tabular}{lrrrrrr}
        \toprule
        & Density & Assortativity & Transitivity & Reciprocity & Avg.\ distance & Mean strength \\
        \midrule
        Density
        & \textbf{-0.311} & -0.026 & -0.091 & \textbf{-0.165} & 0.103 & \textbf{0.818} \\
        Assortativity
        & 0.307 & -0.051 & -0.211 & \textbf{-0.278} & 0.116 & -0.117 \\
        Transitivity
        & \textbf{-0.317} & \textbf{0.250} & \textbf{-0.254} & -0.128 & -0.009 & \textbf{0.990} \\
        Reciprocity
        & -0.303 & -0.070 & -0.037 & \textbf{-0.275} & -0.167 & 0.054 \\
        Avg.\ distance
        & 0.193 & 0.180 & -0.110 & \textbf{0.315} & -0.050 & -0.169 \\
        Mean strength
        & 0.016 & -0.024 & \textbf{-0.160} & 0.063 & \textbf{0.190} & \textbf{0.933} \\
        \bottomrule
    \end{tabular}
    \caption{Estimated statistic-mode coupling matrix $\widehat{\mathbf{A}}_2$ for the MAR(1) model. Columns correspond to source structural characteristics at time $t-1$ and rows to recipient characteristics at time $t$. Bold entries are statistically distinguishable from zero at the nominal $5\%$ level.}
    \label{tab:A2}
\end{table}

Mean strength, by contrast, exhibits the largest statistic-mode coefficients. Its autoregressive entry is $0.933$, while its couplings with subsequent density and transitivity are $0.818$ and $0.990$, respectively. Interaction intensity therefore provides the strongest temporal signal in the statistic mode: fluctuations in mean strength are closely linked to subsequent changes in the prevalence and local closure of international ties. Density displays a different pattern, with negative coefficients for its own subsequent value and for transitivity. The estimated statistic-mode structure therefore distinguishes between the number of active ties and their intensity. Dense configurations do not display the same persistence as configurations characterized by strong interactions, whereas mean strength is associated with pronounced structural carryover.

Taken together, Tables~\ref{tab:A1} and \ref{tab:A2} reveal two distinct dimensions of the dynamics. The \textit{Verbal-} layer is the most prominent source of cross-layer coupling, while mean strength provides the strongest channel of cross-statistic persistence. The separable representation is particularly informative in this respect because these two mechanisms would be embedded within a single $24\times24$ coefficient matrix under an unrestricted VAR formulation.

\begin{figure}[!b]
    \centering
    \includegraphics[width=\linewidth]{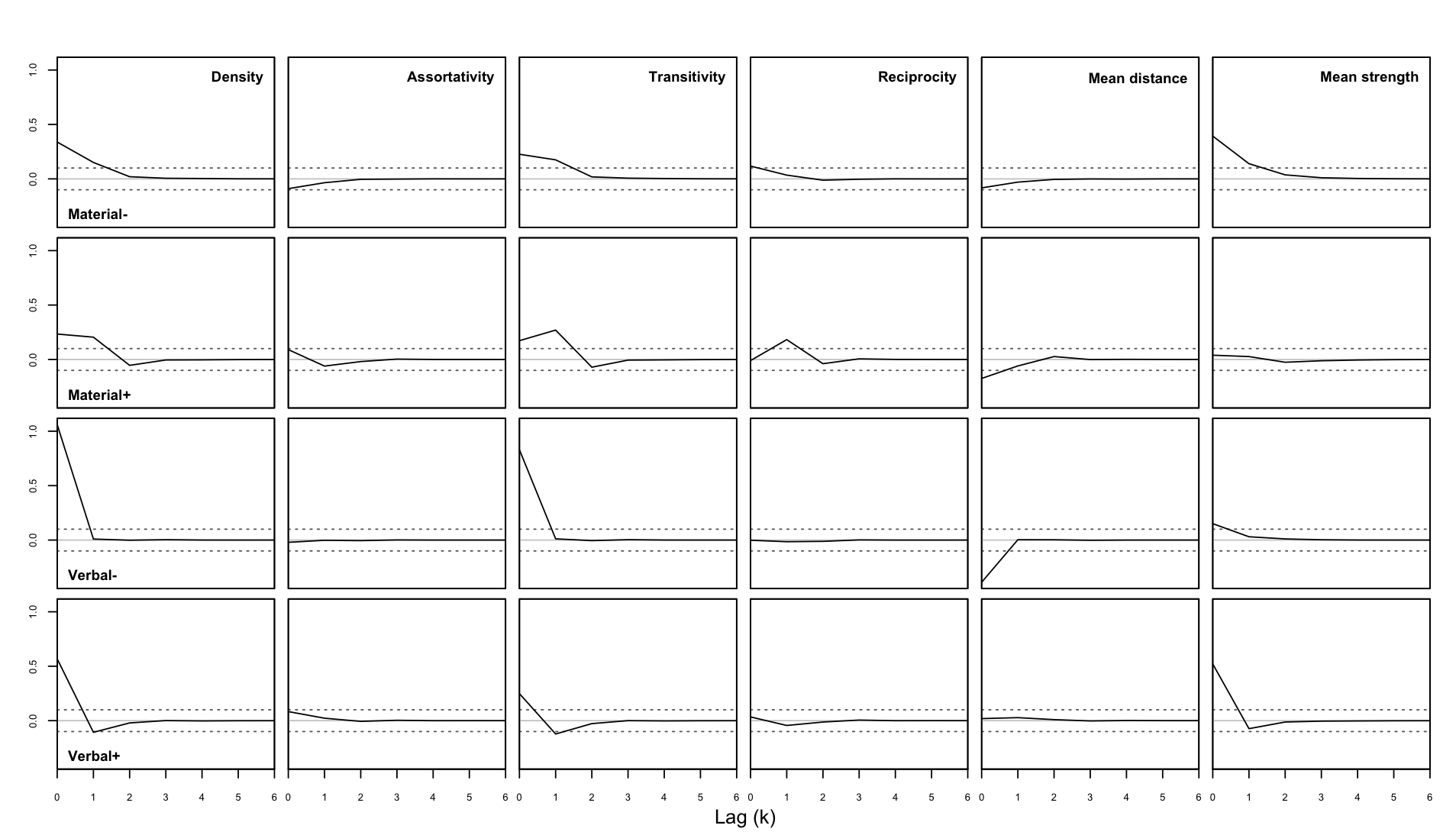}
    \caption{Orthogonalized impulse-response functions following a one-standard-deviation innovation to the density of the \textit{Verbal-} layer.}
    \label{fig:IR_densidad}
\end{figure}

\subsection{Propagation of structural innovations}

The coefficient matrices describe one-step dependence, but the dynamic consequences of a structural innovation may extend across multiple layers, statistics, and time points. We therefore examine orthogonalized impulse-response functions generated from the fitted MAR(1) process. Because these responses are computed from the complete transition matrix $\widehat{\mathbf{A}}_2\otimes\widehat{\mathbf{A}}_1$, they combine layer-mode and statistic-mode dependence and provide a more direct characterization of the propagation of specific structural shocks. The responses are descriptive features of the fitted stochastic system and should not be interpreted as causal effects of international events.

Orthogonalization is based on a Cholesky factorization of the estimated innovation covariance matrix. Consistent with the column-wise vectorization of $\mathbf{X}_t$, the variables are ordered first by structural characteristic and then by layer. The structural characteristics are ordered as density, assortativity, transitivity, reciprocity, average geodesic distance, and mean strength, and within each characteristic the layers are ordered as \textit{Material-}, \textit{Material+}, \textit{Verbal-}, and \textit{Verbal+}.

We focus first on a one-standard-deviation innovation to the density of the \textit{Verbal-} layer. Figure~\ref{fig:IR_densidad} shows that the disturbance propagates beyond its originating layer and is accompanied by changes in interaction intensity elsewhere in the multilayer system. The dominant responses are concentrated at short horizons and largely dissipate within one or two months. By contrast, assortativity and reciprocity exhibit little persistent response. This pattern indicates that a sudden increase in the prevalence of negative verbal interactions is associated with a broad but transient reorganization of the structural state rather than a sustained alteration of its longer-lasting mixing or reciprocity patterns.

\begin{figure}[!b]
    \centering
    \includegraphics[width=\linewidth]{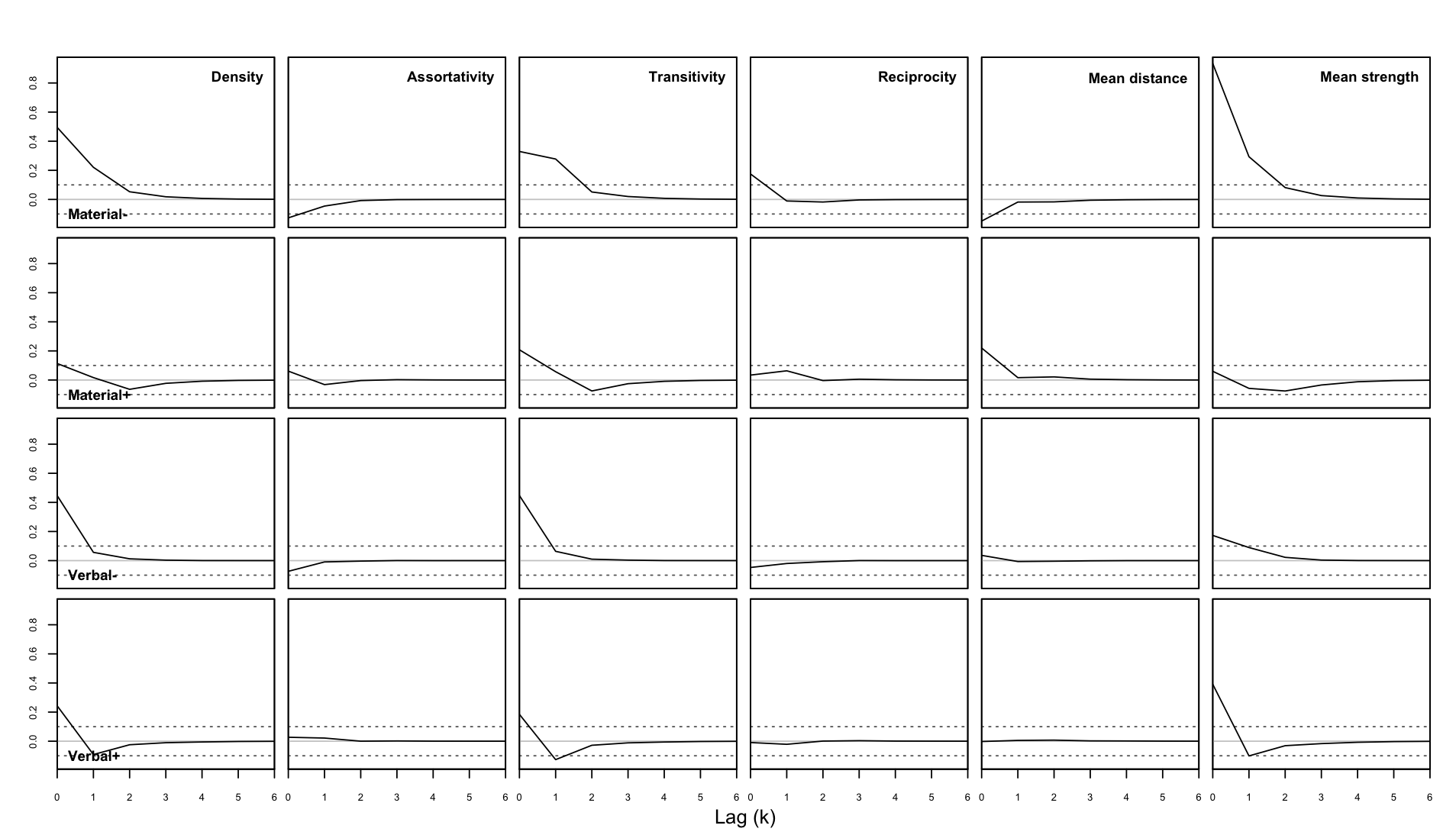}
    \caption{Orthogonalized impulse-response functions following a one-standard-deviation innovation to the mean strength of the \textit{Material-} layer.}
    \label{fig:IR_fuerza}
\end{figure}

The second experiment considers a one-standard-deviation innovation to mean strength in the \textit{Material-} layer. The responses in Figure~\ref{fig:IR_fuerza} differ markedly from those generated by the \textit{Verbal-} density shock. Increased interaction intensity in material conflict is followed by persistent reinforcement of the negative-material layer, including higher density, greater transitivity, and sustained interaction strength. The disturbance is also associated with a short-run deterioration in the structural state of verbal cooperation, concentrated approximately within the first two subsequent months, while the response of positive material interactions is comparatively limited.

The two impulse-response experiments therefore reveal a substantive distinction between the \emph{reach} and the \emph{persistence} of structural innovations. A density innovation in verbal conflict propagates broadly but dissipates rapidly, whereas an innovation in the intensity of material conflict generates a more persistent response and is accompanied by a temporary weakening of verbal cooperation. This difference reinforces the role of mean strength as the principal carrier of temporal persistence identified in $\widehat{\mathbf{A}}_2$, while also confirming that the dominant \textit{Verbal-} layer-mode coupling does not necessarily imply long-lasting responses to every type of verbal-conflict innovation.

\subsection{Model adequacy}

\begin{figure}[!b]
    \centering
    \includegraphics[width=0.92\linewidth]{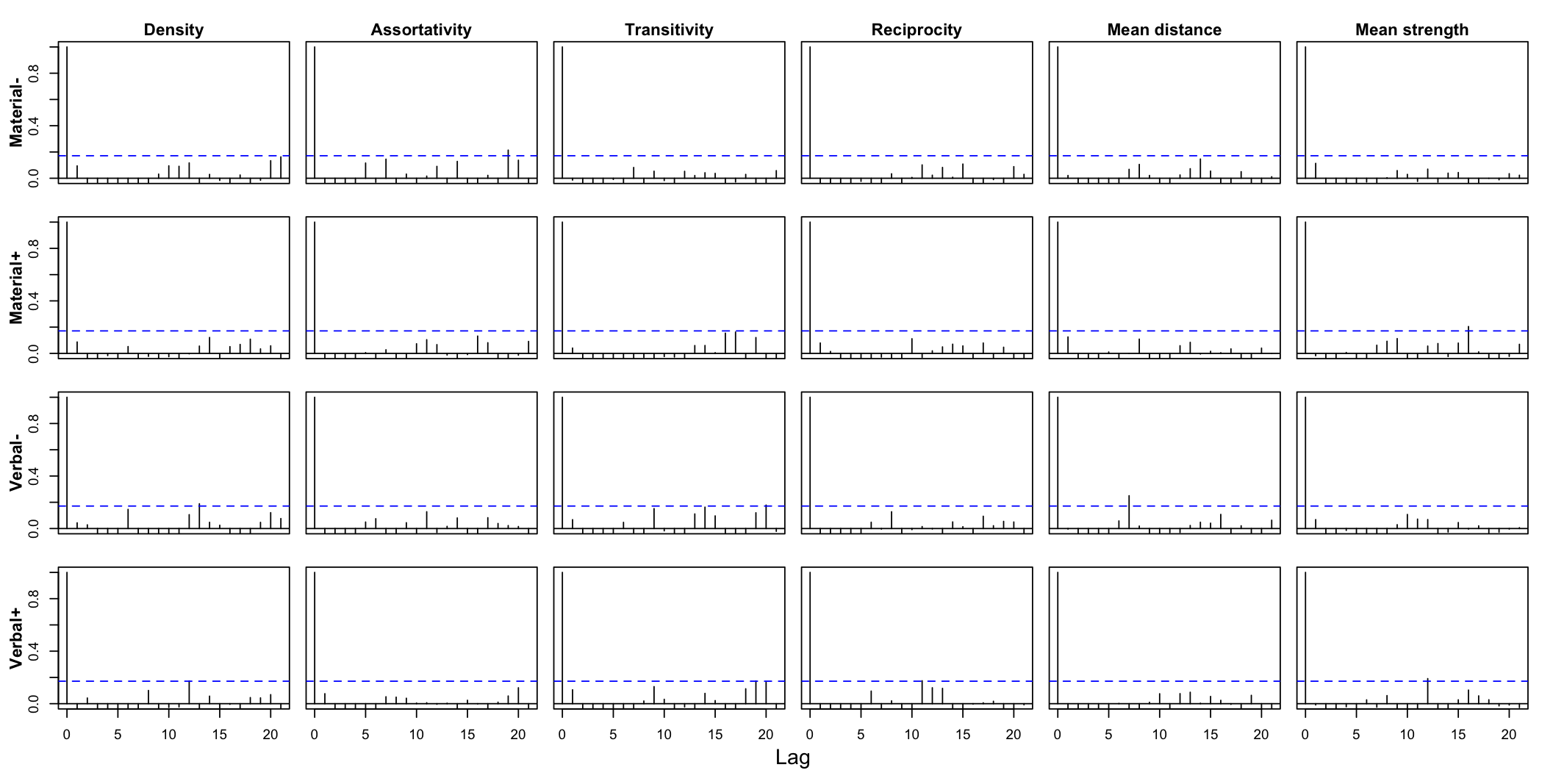}
    \caption{Autocorrelation functions for the $24$ residual series from the fitted MAR(1) model.}
    \label{fig:res_acf}
\end{figure}

\begin{figure}[!t]
    \centering
    \includegraphics[width=0.92\linewidth]{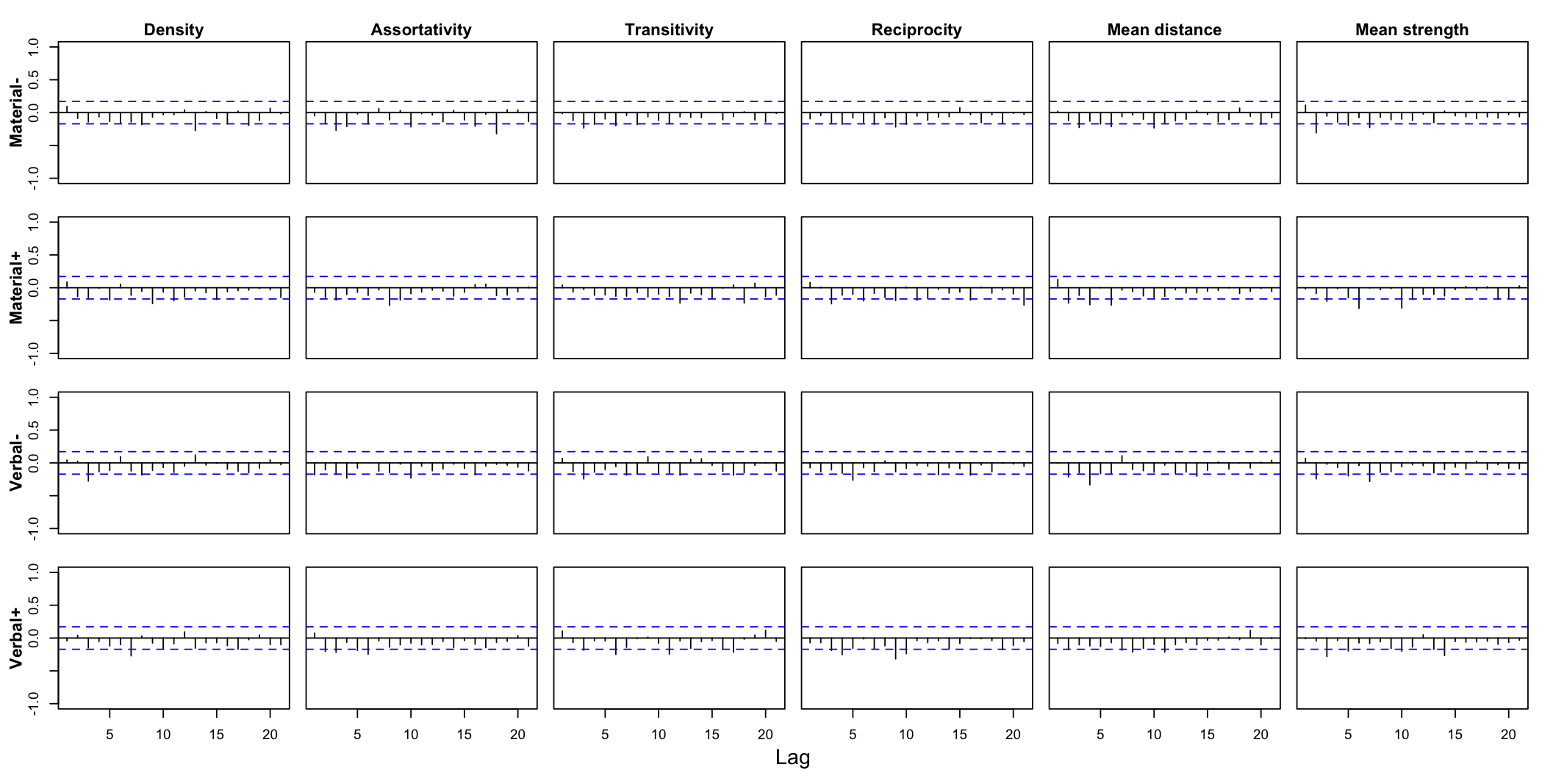}
    \caption{Partial autocorrelation functions for the $24$ residual series from the fitted MAR(1) model.}
    \label{fig:res_pacf}
\end{figure}

Model adequacy is assessed through graphical and formal diagnostics of the $24$ residual component series. Figures~\ref{fig:res_acf} and \ref{fig:res_pacf} display their autocorrelation and partial autocorrelation functions. Most correlations remain within the corresponding sampling bounds, with relatively few isolated departures, indicating that the fitted MAR(1) model removes most of the systematic short-run temporal dependence.

\begin{figure}[!htb]
    \centering
    \includegraphics[width=0.92\linewidth]{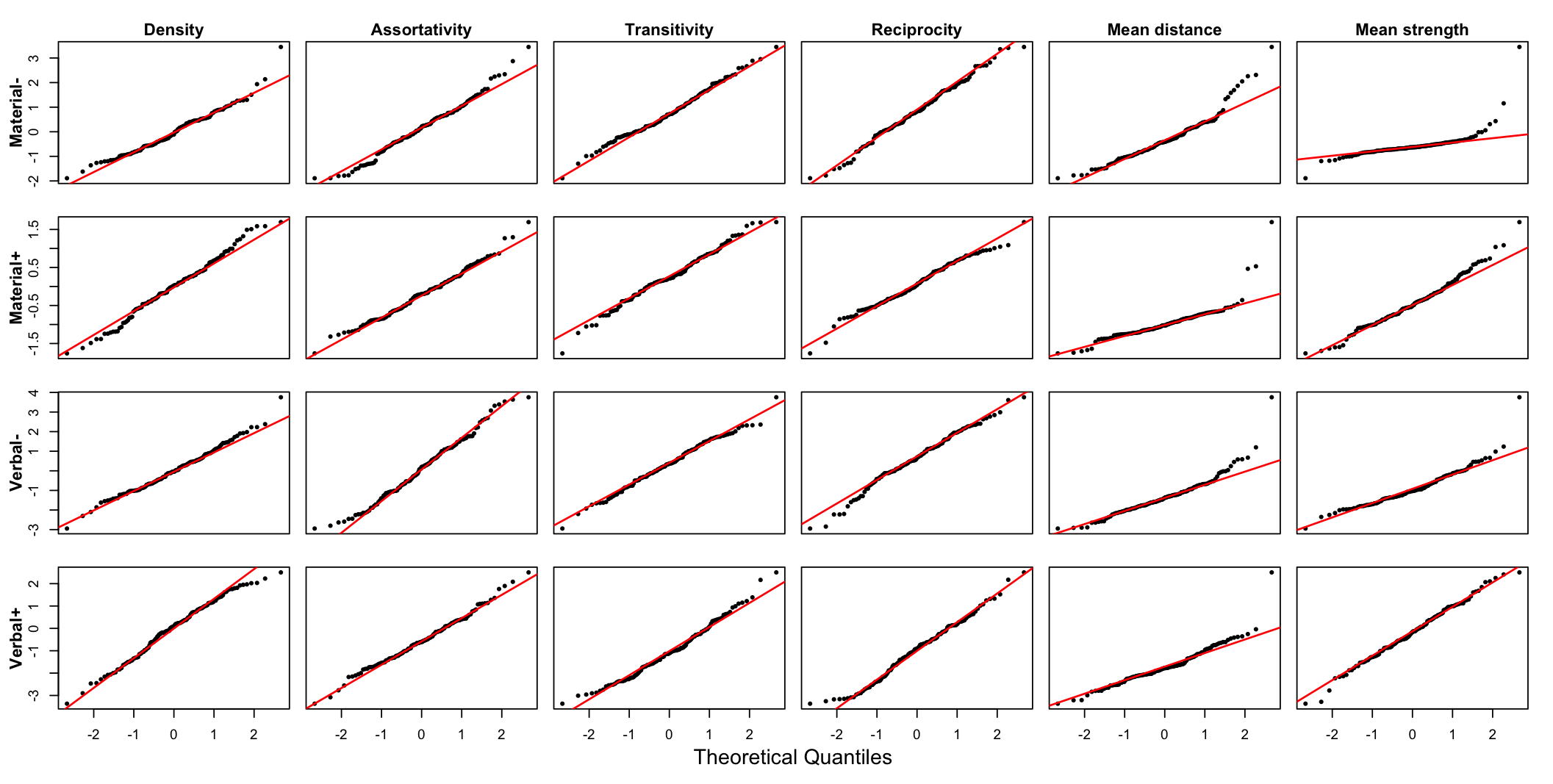}
    \caption{Normal Q--Q plots for the $24$ residual series from the fitted MAR(1) model.}
    \label{fig:res_qq}
\end{figure}

The normal Q--Q plots in Figure~\ref{fig:res_qq} show approximate Gaussian behavior for most residual components, with departures concentrated mainly in the tails of some series. This diagnostic is used descriptively: Gaussian innovations are not required for consistency of the least-squares estimator under the regularity conditions underlying the MAR estimation theory, although substantial deviations may still indicate unmodeled features of the innovation process.

Remaining serial dependence is examined formally using the Ljung--Box test \citep{Ljung1978} for each residual component. Componentwise diagnostics are preferred here to a conventional multivariate portmanteau test such as that of \citet{Hosking1980}. With $N=24$ jointly modeled series and only $T=132$ time points, the dimension is large relative to the available temporal information, a setting in which the usual asymptotic approximation underlying multivariate portmanteau statistics can be unreliable \citep{Lutkepohl2005}. The componentwise tests are therefore used to identify residual temporal structure without relying on a potentially inaccurate high-dimensional joint approximation.

Because $24$ hypotheses are evaluated simultaneously, their $p$-values are adjusted using the Benjamini--Hochberg false discovery rate procedure \citep{Benjamini1995}. After adjustment, only $3$ of the $24$ residual series retain evidence of serial dependence at the $5\%$ level. Equivalently, $21$ of the $24$ components, or $87.5\%$, show no detectable remaining autocorrelation under this diagnostic. Because the residual components are themselves cross-sectionally related, these tests are interpreted as multiplicity-adjusted componentwise diagnostics rather than as a formal proof of joint white noise.

The graphical diagnostics, componentwise Ljung--Box results, and BIC comparison across lag orders provide mutually consistent evidence that the parsimonious MAR(1) specification captures most of the short-run temporal structure in the matrix-valued series. Residual dependence is not eliminated completely, but the remaining departures are limited relative to the $24$ simultaneously modeled components and do not dominate the fitted dynamics.

Overall, the empirical analysis reveals a structured and strongly asymmetric system of international-interaction dynamics. Negative verbal interactions constitute the most prominent source of cross-layer coupling, whereas interaction intensity, summarized by mean strength, provides the strongest channel of persistence across structural characteristics. The impulse-response analysis further shows that these mechanisms operate on different temporal scales: innovations in verbal-conflict density produce broad but rapidly dissipating structural responses, while innovations in the intensity of material conflict generate more persistent reconfiguration and are accompanied by a temporary weakening of verbal cooperation. These findings illustrate the principal inferential advantage of the structural-state representation: it separates the relational domain from which temporal variation originates, the network characteristics through which that variation is transmitted, and the horizon over which the resulting structural adjustments persist.


\section{Simulation Study}

The empirical analysis in Section~3 relies on a bilinear MAR representation for the temporal evolution of structural summaries of a dynamic multilayer network. Because the underlying evolution of a network need not itself follow a bilinear process, we conduct a simulation study to evaluate the extent to which the proposed structural-state representation can recover known cross-layer temporal dependence when the observed networks arise from a substantially more complex nonlinear mechanism. The simulation is therefore not designed to reproduce the MAR model by construction. Instead, it evaluates whether a MAR model fitted to global network statistics can recover the dominant dependence structure after edge formation, thresholding, triadic closure, and reciprocity have transformed the underlying dynamics.

\subsection{Data-generating mechanism}

We generate directed dynamic multilayer networks with four layers observed over $T$ time points. For each ordered pair of nodes $(i,j)$, let
\[
\mathbf{z}_{ij,t}
=
\left(
z_{ij,t}^{(1)},\ldots,z_{ij,t}^{(4)}
\right)^\top
\]
denote the vector of latent interaction intensities across layers at time $t$. The temporal evolution of these intensities follows a first-order vector autoregressive mechanism driven by a prespecified $4\times4$ coupling matrix $\mathbf{C}$,
\[
\mathbf{z}_{ij,t}
=
\mathbf{C}\mathbf{z}_{ij,t-1}
+
\boldsymbol{\varepsilon}_{ij,t},
\]
where the innovations are Gaussian with standard deviation $\sigma_z$. Each resulting component is truncated to $[0,1]$ and subsequently thresholded. An edge is considered active whenever the corresponding latent intensity exceeds a threshold $\tau$. The coupling matrix therefore determines the true temporal dependence among layers at the latent edge level.

This autoregressive component is deliberately supplemented with mechanisms that induce network dependence beyond dyad-specific persistence. In particular, an additional autoregressive process determines a time-varying triangulation propensity for each layer. This mechanism increases the probability of forming an edge between two actors that share a common neighbor, thereby inducing endogenous transitivity. A partial-reciprocity mechanism subsequently converts a random fraction of directed dyads into reciprocal relationships. Consequently, although cross-layer persistence is introduced through a linear latent process, the observed network sequence is generated through nonlinear transformations involving truncation, thresholding, triadic closure, and reciprocity.

This distinction is important for the purpose of the experiment. The MAR model is not fitted to the latent quantities $\mathbf{z}_{ij,t}$ and is not expected to reproduce the numerical entries of $\mathbf{C}$. Instead, each simulated network is reduced to the same structural-state representation used in the ICEWS application. For every layer and time point, we compute density, assortativity, transitivity, reciprocity, average geodesic distance, and mean strength. After standardization, these quantities form a $4\times6$ matrix-valued time series to which a MAR(1) model is fitted. The simulation therefore evaluates whether the \emph{support} of the underlying cross-layer dependence, that is, which layer-to-layer temporal connections are present or absent, remains detectable after the nonlinear network-generating mechanism and the subsequent structural aggregation.

This controlled construction is used because it allows the cross-layer dependence structure to be specified explicitly and independently of the additional network mechanisms. The essential feature of the design is therefore that the true cross-layer coupling structure is known, while the resulting structural-state process does not follow a MAR model by construction.

\subsection{Cross-layer coupling structure}

The baseline coupling matrix is
\[
\mathbf{C}
=
\begin{pmatrix}
0.6 & 0   & 0   & 0 \\
0   & 0   & 0   & 0 \\
0.5 & 0   & 0.3 & 0 \\
0   & 0   & 0.4 & 0.5
\end{pmatrix}.
\]

Under the convention that columns correspond to source layers at time $t-1$ and rows to recipient layers at time $t$, this specification generates a sparse and directionally asymmetric dependence structure. Layer~1 exhibits autoregressive persistence of $0.6$. Layer~3 depends both on its own preceding state, with coefficient $0.3$, and on Layer~1, with coefficient $0.5$. Layer~4 has autoregressive persistence of $0.5$ and additionally depends on Layer~3 through a coefficient of $0.4$. The resulting cross-layer pathway includes the directed chain $1\rightarrow3\rightarrow4$. Layer~2 is dynamically isolated under the baseline specification, having neither autoregressive persistence nor incoming cross-layer dependence.

The sparse structure of $\mathbf{C}$ allows the simulation to assess two distinct properties of the fitted MAR representation. Nonzero entries provide known layer-level dependencies that should ideally be detected, whereas zero entries provide a substantially larger collection of absent relationships against which false-positive behavior can be evaluated.

\subsection{Simulation design}

The baseline configuration uses $n=25$ nodes, $T=132$ time points, innovation standard deviation $\sigma_z=0.05$, and edge threshold $\tau=0.15$. Nine scenarios are considered to evaluate robustness with respect to the strength and location of cross-layer dependence, stochastic variability, network size, temporal information, and the threshold governing edge realization. Table~\ref{tab:simulation_scenarios} summarizes the design.

\begin{table}[!tbp]
    \centering
    \small
    \begin{tabular}{cl}
        \toprule
        Scenario & Modification relative to the baseline configuration \\
        \midrule
        1 & Baseline coupling matrix and baseline simulation parameters. \\
        2 & Null coupling matrix, $\mathbf{C}=\mathbf{0}$. \\
        3 & All nonzero entries of $\mathbf{C}$ reduced by one half. \\
        4 & Increased edge-level noise, $\sigma_z=0.15$. \\
        5 & Smaller network, $n=12$. \\
        6 & Larger network, $n=50$. \\
        7 & Shorter time series, $T=60$. \\
        8 & Cross-layer dependence introduced also through the triangulation process. \\
        9 & Higher edge-activation threshold, $\tau=0.25$. \\
        \bottomrule
    \end{tabular}
    \caption{Simulation scenarios. Unless otherwise indicated, $n=25$, $T=132$, $\sigma_z=0.05$, $\tau=0.15$, and the baseline coupling matrix $\mathbf{C}$ are used.}
    \label{tab:simulation_scenarios}
\end{table}

The scenarios isolate different sources of difficulty. Scenario~2 provides a direct assessment of false-positive behavior because no true temporal coupling is present. Scenario~3 evaluates detection under weaker signals, while Scenario~4 increases stochastic variation relative to the systematic layer dependence. Scenarios~5 and~6 examine the effect of network size. Scenario~7 reduces the amount of temporal information available for MAR estimation. Scenario~8 is particularly informative because cross-layer dependence is introduced through the triangulation mechanism in addition to the edge-weight process, thereby assessing whether dependence remains detectable when it operates partly through higher-order network structure rather than exclusively through direct edge persistence. Finally, Scenario~9 changes network sparsity by increasing the threshold required for an interaction to be observed.

Each scenario is independently replicated $B=40$ times. For every replicate, the six standardized structural characteristics are assembled into the matrix-valued series and a MAR(1) model is estimated by least squares. The estimated layer-mode matrix is then compared with the known support of the data-generating coupling structure.

We summarize support recovery using sensitivity and specificity. Sensitivity is the proportion of truly nonzero layer couplings identified as statistically distinguishable from zero at the nominal $5\%$ level, whereas specificity is the proportion of true zero couplings that are not declared significant. Scenario~2 has no sensitivity value because its coupling matrix contains no nonzero entries. Importantly, a value of $0.95$ is not a theoretical benchmark for sensitivity. For specificity, however, approximately $0.95$ provides a natural nominal reference under correctly calibrated individual tests conducted at level $0.05$. Thus, the two measures address distinct questions: sensitivity quantifies the ability to recover genuine dependence, whereas specificity measures protection against spurious cross-layer connections.

\subsection{Recovery of cross-layer dependence}

\begin{figure}[!b]
    \centering
    \includegraphics[width=0.9\linewidth]{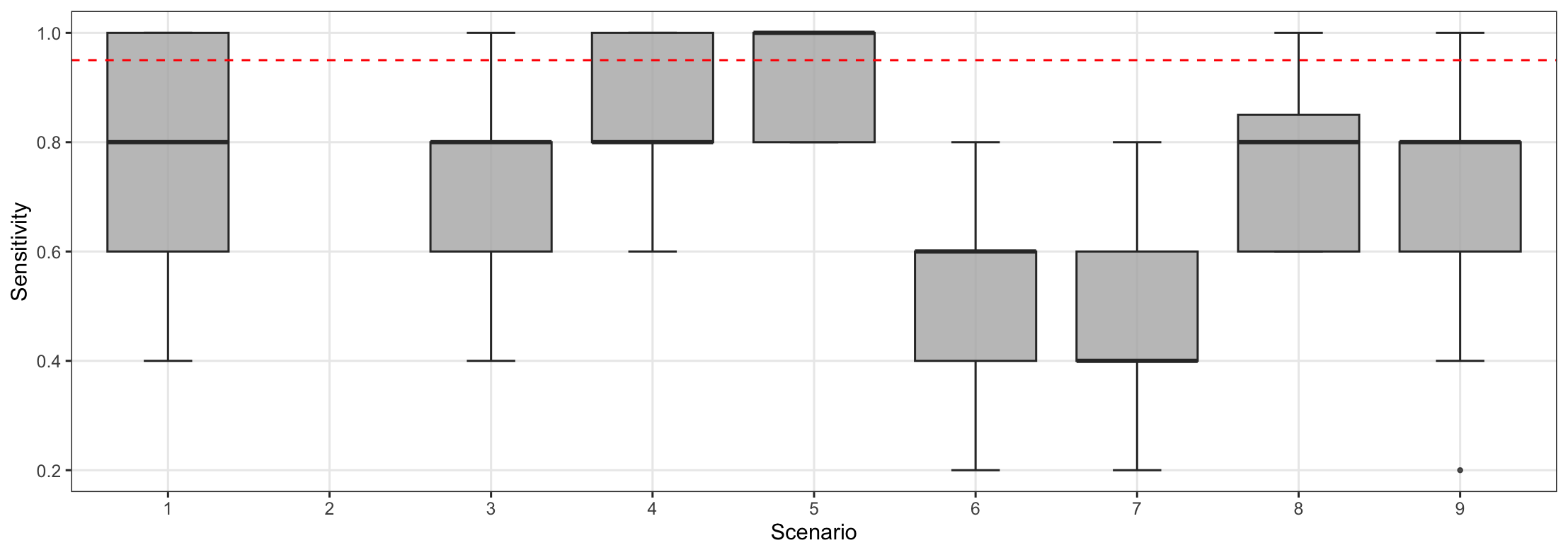}
    \caption{Sensitivity for recovery of true cross-layer dependencies across the simulation scenarios. Each boxplot summarizes $B=40$ independent replications. Scenario~2 is omitted because the corresponding coupling matrix contains no nonzero entries.}
    \label{fig:sensib}
\end{figure}

Figure~\ref{fig:sensib} displays the distribution of sensitivity across the $40$ replications of each applicable scenario. The principal result is that the MAR representation retains substantial ability to identify genuine layer-level dependence despite the nonlinear transformation separating the latent edge dynamics from the observed structural summaries. Sensitivity is concentrated near the upper end of its range in most scenarios, indicating that true temporal couplings are rarely missed. This is noteworthy because estimation is performed exclusively on global network characteristics rather than on the edge-level variables through which the dependence was originally introduced.

The behavior across scenarios also clarifies when support recovery becomes more difficult. Scenario~5, based on networks with $n=12$ nodes, produces the highest detection of the true couplings (median sensitivity = $1$). In contrast, the clearest deterioration occurs in Scenario~7, where the time series is shortened from $132$ to $60$ observations (median sensitivity = $0.4$). This result is consistent with the fact that the effective information for estimating the autoregressive coefficients is primarily temporal rather than determined simply by the number of dyads in each network. Reduced $T$ therefore translates directly into greater uncertainty in the estimated MAR dynamics.

Sensitivity also decreases in Scenario~6 when the network size is increased to $n=50$. This finding is initially less intuitive because larger networks contain more dyadic information. However, the MAR model is fitted to six global summaries per layer rather than to the individual dyads. Increasing $n$ does not increase the number of observations in the matrix-valued time series and may instead alter the variability and concentration of the structural summaries produced by the underlying network process. The result therefore emphasizes that the effective sample size for MAR estimation remains governed by $T$, while network size affects the stochastic properties of the aggregated structural state.

The remaining perturbations, including weaker coupling, increased edge-level noise, changes in sparsity, and the introduction of dependence through triangulation, do not eliminate the ability of the MAR model to recover the dominant layer-level structure. In particular, Scenario~8 provides evidence that the proposed representation can detect cross-layer dependence even when part of that dependence propagates through higher-order network organization rather than exclusively through direct edge persistence.

\subsection{False-positive behavior}

\begin{figure}[!tbp]
    \centering
    \includegraphics[width=0.9\linewidth]{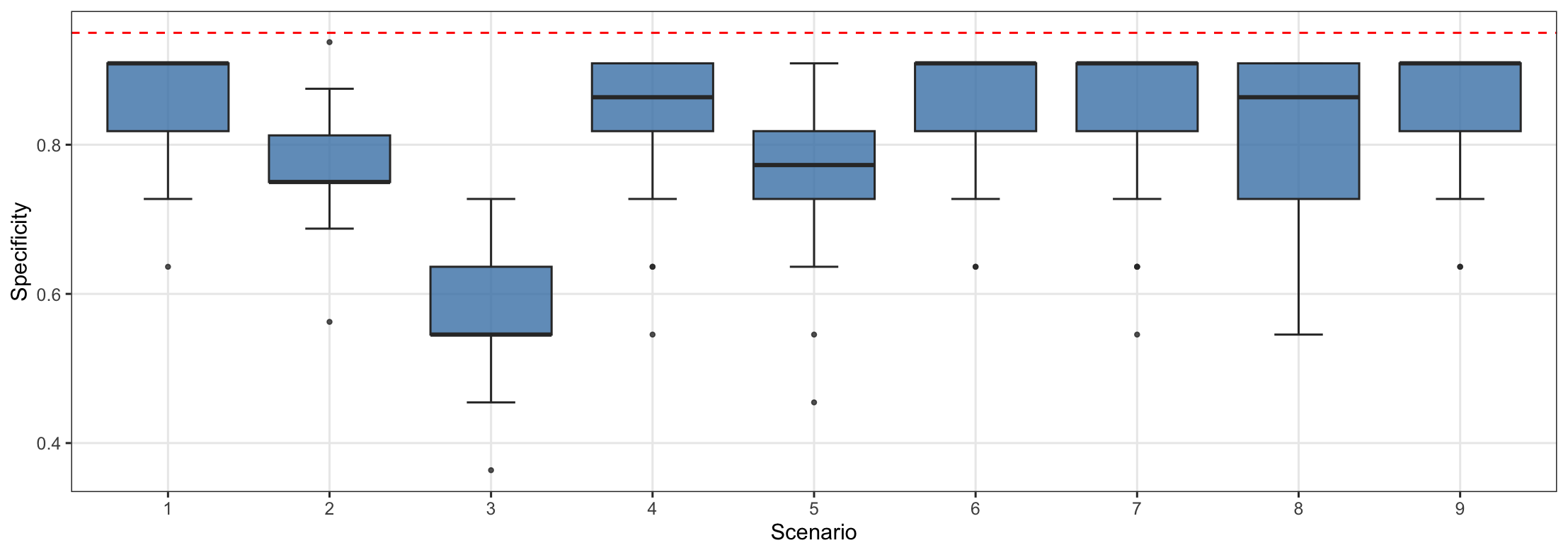}
    \caption{Specificity for recovery of absent cross-layer dependencies across the simulation scenarios. Each boxplot summarizes $B=40$ independent replications.}
    \label{fig:espec}
\end{figure}

Sensitivity alone does not establish reliable recovery of the dependence structure. Figure~\ref{fig:espec} reports specificity and reveals a substantially different aspect of model performance. Specificity varies considerably across scenarios rather than clustering uniformly below the nominal threshold. Scenarios~1, 6, 7, and~9 achieve the highest median specificity ($0.91$), still below the nominal reference value of $0.95$ that would correspond to correctly calibrated $5\%$ tests for individual null coefficients. Specificity is markedly lower, however, in Scenario~3 (median = $0.545$) and, to a lesser extent, in Scenarios~2 and~5 (medians of $0.750$ and $0.773$, respectively). The fitted MAR model therefore tends to recover true connections but also identifies a non-negligible number of layer-level associations that are absent from the data-generating coupling matrix, with the extent of this behavior depending markedly on the scenario considered.

This asymmetry between sensitivity and specificity is the central result of the simulation study. The structural-state representation is effective at preserving strong temporal signals generated at the network level, even after nonlinear edge formation and aggregation into global statistics. It is substantially less reliable, however, for exact recovery of the zeros in the underlying coupling structure. In other words, the bilinear approximation tends to favor detection over sparsity.

Several features of the experiment may contribute to this behavior. The MAR model operates on nonlinear global summaries of the generated networks rather than on the latent edge process itself. Thresholding, transitivity, and reciprocity can transmit dependence between structural characteristics even when a direct entry of $\mathbf{C}$ is zero. Consequently, an estimated MAR coupling need not correspond one-to-one with an edge-level coupling coefficient. Some apparent false positives relative to the support of $\mathbf{C}$ may therefore represent indirect dependence induced by the network-generating mechanism rather than purely numerical estimation error. This observation reinforces the distinction between recovering the dominant structural pathways of the observed network process and exactly reconstructing the latent coupling matrix.

Scenario~6, corresponding to the largest networks, attains specificity comparable to the best-performing settings in the study (median = $0.909$, tied with Scenarios~1, 7, and~9). This contrasts with its markedly weaker sensitivity (median = $0.60$) and illustrates a genuine detection trade-off: configurations that reduce the number of spurious connections need not maximize the probability of identifying every true coupling.

\subsection{Implications of the simulation study}

Taken together, the simulations provide qualified support for the use of a bilinear MAR approximation for structural-state dynamics. Across several departures from the baseline configuration, including increased noise, weaker dependence, changes in sparsity, and cross-layer dependence acting through higher-order network structure, the model generally preserves high sensitivity to genuine temporal coupling. This ability degrades, however, under configurations that reduce the effective information available for estimation, namely shorter temporal sequences and, to a somewhat lesser extent, larger networks, both of which lowered sensitivity substantially relative to the baseline. Thus, while the principal cross-layer pathways can remain visible at the level of global network statistics even when the observed network process is generated through nonlinear mechanisms, this visibility is conditional on the temporal series being sufficiently long relative to the dimensionality of the structural state being tracked.

The study simultaneously identifies an important limitation. Specificity was in no scenario as high as would be expected from perfectly calibrated coefficient-level testing, and its magnitude varied considerably across configurations, ranging from values close to the best attainable in the study under baseline-like conditions to markedly lower values when the true coupling was weak. This implies that individual nonzero MAR coefficients should not be interpreted as definitive evidence of a corresponding direct edge-level dependence, and that this risk is not uniform but increases specifically when the underlying dependence is weak relative to the noise in the structural summaries. The method is therefore better suited to identifying dominant channels of structural propagation than to exact recovery of a sparse latent coupling graph.

This distinction is directly relevant to the ICEWS analysis. The empirical findings concerning \textit{Verbal-} and mean strength should be interpreted as evidence about prominent dependencies in the evolution of the observed structural state, rather than as a claim that the estimated MAR coefficients recover a unique underlying mechanism of international interaction. The simulations nevertheless show that strong and systematic cross-layer signals can survive substantial nonlinear transformation and remain detectable through the proposed matrix-valued representation. In this sense, the experiment supports the central methodological premise of the paper: a parsimonious MAR model can provide an informative approximation to the macroscopic dynamics of a complex multilayer network even when the underlying edge-level evolution is not itself bilinear.


\section{Discussion}

This paper develops a structural-state approach for analyzing the temporal evolution of dynamic multilayer networks. Each layer is summarized through global network statistics, yielding a matrix-valued time series that is modeled using a matrix autoregressive specification. This representation preserves the distinction between relational layers and structural characteristics while providing a parsimonious description of macroscopic network dynamics.

The ICEWS application reveals a strongly asymmetric pattern of cross-layer dependence. Negative verbal interactions emerge as the most prominent source of subsequent structural variation, with their lagged configuration associated with changes in both material-interaction layers. In contrast, the structural state of verbal conflict is not significantly explained by the preceding configuration of the other layers. Positive verbal interactions exhibit much weaker propagation. These results identify negative verbal relations as a distinctive channel of short-run structural dependence, although the associations should not be interpreted causally.

At the level of structural characteristics, mean strength provides the strongest temporal signal. It exhibits substantial persistence and is strongly associated with subsequent density and transitivity, indicating that interaction intensity generates greater structural momentum than the mere prevalence of ties. Density instead displays short-run mean reversion, while reciprocity is associated with subsequent reductions in connectivity and greater average separation. The impulse-response analysis reinforces these findings. Innovations in verbal-conflict density propagate broadly but dissipate within a few periods, whereas innovations in the mean strength of material conflict generate more persistent responses and are accompanied by a temporary weakening of verbal cooperation.

The simulation study provides qualified support for the bilinear approximation. Even though the generated networks arise from nonlinear mechanisms involving thresholding, triadic closure, and reciprocity, the MAR representation generally achieves high sensitivity for detecting genuine cross-layer dependence. However, specificity is substantially lower, indicating that some estimated dependencies do not correspond to direct nonzero entries of the latent coupling matrix. Accordingly, the method is better suited to identifying dominant channels of structural propagation than to exact recovery of an underlying sparse dependence graph.

The approach also has important limitations. Structural summaries necessarily discard dyad-specific information, and the bilinear MAR specification imposes a separable dependence structure that may be restrictive for more complex interactions. In addition, the empirical analysis relies on detrending and an approximately stationary short-run process. Future work could consider higher-order, sparse, time-varying, nonstationary, or more general matrix and tensor autoregressive models \citep{Review}, as well as multiscale formulations that connect global structural dynamics with underlying edge-level mechanisms.

Overall, the proposed framework provides a concise and interpretable way to separate where structural change originates, which network characteristics transmit it, and how long its effects persist. In the ICEWS system, negative verbal interactions emerge as the dominant cross-layer signal, while mean interaction strength is the main carrier of temporal persistence.

\section*{Generative-AI assistance disclosure}

Generative AI was used to assist drafting, structural editing, and LaTeX preparation. The authors are responsible for verifying the sources, arguments, mathematical statements, and final wording before submission. No confidential data were used in preparing this manuscript.

\section*{Declarations}

\noindent\textbf{Funding.} No external funding was received for the preparation of this manuscript.

\noindent\textbf{Conflict of interest.} The authors declare no conflict of interest.

\noindent\textbf{Data availability.} The ICEWS data analyzed in this study are publicly available through the Harvard Dataverse at:\\
\url{https://dataverse.harvard.edu/dataverse/icews}.

\noindent\textbf{Code availability.} The code used to reproduce the analyses presented in this study is publicly available on GitHub at:\\
\url{https://github.com/CamilaPinzon19/Multilayer-dynamic-network-analysis}.

\bibliographystyle{apalike}
\bibliography{references.bib}
\end{document}